\documentclass[slac_one]{revtex4}
\usepackage{graphicx}
\usepackage{fancyhdr}
\newcommand{\beq}{\begin{equation}}
\newcommand{\eeq}{\end{equation}}
\newcommand{\bea}{\begin{eqnarray}}
\newcommand{\eea}{\end{eqnarray}}
\newcommand{\Fig}[1]{Fig.\,\ref{#1}}

\newcommand{\Eq}[1]{Eq.\,(\ref{#1})}

\newcommand{\Tab}[1]{Tab.\,\ref{#1}}

\newcommand{\f}{\frac}

\newcommand{\sw}{s_{\scriptscriptstyle W}}
\newcommand{\cw}{c_{\scriptscriptstyle W}}

\newcommand{\MZ}{M_{\scriptscriptstyle Z}}

\newcommand{\mb}{m_b}
\newcommand{\mt}{m_t}

\newcommand{\GF}{G_F}

\newcommand{\GeV}{{\rm GeV}}
\newcommand{\TeV}{{\rm TeV}}

\newcommand{\re}{{\rm Re}}

\def\unit{\leavevmode\hbox{\small1\kern-3.6pt\normalsize1}}

\newcommand{\sL}{{\scalebox{0.6}{$L$}}}

\newcommand{\Kpnn}{K^+ \to \pi^+ \nu \bar{\nu}}
\newcommand{\KLnn}{K_L \to \pi^0 \nu \bar{\nu}}

\newcommand{\KLmm}{K_L \to \mu^+ \mu^-}

\newcommand{\BXsga}{\bar{B} \to X_s \gamma}

\newcommand{\BXsll}{\bar{B} \to X_s l^+ l^-}

\newcommand{\BXdnn}{\bar{B} \to X_d \nu \bar{\nu}}
\newcommand{\BXsnn}{\bar{B} \to X_s \nu \bar{\nu}}

\newcommand{\Bdmm}{B_d \to \mu^+ \mu^-}
\newcommand{\Bsmm}{B_s \to \mu^+ \mu^-}

\newcommand{\BRKp}{{\cal B} (\Kpnn)}
\newcommand{\BRKL}{{\cal B} (\KLnn)}
\newcommand{\BRKm}{{\cal B} (\KLmm)_{\rm SD}}
\newcommand{\BRXd}{{\cal B} (\BXdnn)}
\newcommand{\BRXs}{{\cal B} (\BXsnn)}
\newcommand{\BRBd}{{\cal B} (\Bdmm)}
\newcommand{\BRBs}{{\cal B} (\Bsmm)}
\newcommand{\BRga}{{\cal B} (\BXsga)}
\newcommand{\BRll}{{\cal B} (\BXsll)}

\newcommand{\btosgamma}{b \to s \gamma}

\newcommand{\Ztobb}{Z \to b \bar{b}}
\newcommand{\Ztobs}{b \to s Z}

\newcommand{\Ztodjdi}{Z \to d^j \bar{d}^i}
\newcommand{\Zbb}{Z b_\sL \bar{b}_\sL}
\newcommand{\Zdjdi}{Z d^j_\sL \bar{d^i_\sL}}

\newcommand{\C}{C}

\newcommand{\Rb}{R_b^0}
\newcommand{\Ab}{{\cal A}_b}
\newcommand{\AFB}{A_{\rm FB}^{0, b}}

\newcommand{\mysigma}{\hspace{0.4mm} \sigma}
\newcommand{\etal}{{\it et al}.}

\begin{document}


\title{ \boldmath \hspace{5mm} \mbox{Something about $Z$-penguins
    I want to tell}  \unboldmath }

\author{Ulrich~Haisch} 

\affiliation{ Institut f\"ur Physik (THEP), Johannes
  Gutenberg-Universit\"at, D-55099 Mainz, Germany }


\begin{abstract}

\noindent
We stress that in models with constrained minimal flavor violation
large negative corrections to the flavor-changing $Z$-penguin
amplitudes are excluded by the precision measurements of the $\Ztobb$
pseudo observables performed at LEP and SLC. The derived stringent
range for the non-standard contribution to the universal Inami-Lim
function $\C$ leads to tight two-sided limits for the branching ratios
of all $Z$-penguin dominated flavor-changing $K$- and $B$-decays.
\end{abstract}

\maketitle

\thispagestyle{fancy}

\section{INTRODUCTION}
\label{sec:introduction}

The effects of new heavy particles appearing in extensions of the
standard model (SM) can be accounted for at low energies in terms of
effective operators. The unprecedented accuracy reached by the
electroweak (EW) precision measurements performed at the high-energy
colliders LEP and SLC impose stringent constraints on the coefficients
of the operators entering the EW sector. Other severe constraints came
in recent years from the BaBar, Belle, CDF, and D\O\ experiments and
concern extra sources of flavor and CP violation that represent a
generic problem in many beyond the SM (BSM) scenarios. The most
pessimistic but experimentally well supported solution to the flavor
puzzle is to assume that all flavor and CP violation is governed by
the known structure of the SM Yukawa interactions. In these minimal
flavor violating (MFV) \cite{Buras:2000dm, D'Ambrosio:2002ex} models,
correlations between certain flavor diagonal high-energy and flavor
off-diagonal low-energy observables exist since, by construction, BSM
physics couples dominantly to the third generation. To simplify
matters, we restrict ourselves in the following to scenarios that
involve only SM operators, so-called constrained MFV (CMFV)
\cite{Blanke:2006ig} models, and thus consider only left-handed
currents. Correlations between flavor diagonal and off-diagonal
amplitudes, similar to the ones discussed below, might exist in many
beyond-MFV scenarios in which the modification of the flavor structure
is non-universal. One example for such a correlation is provided by
the intimate relation between the $\Ztobs$ and $\Ztobb$ amplitude
\cite{Casagrande:2008hr} present in the original Randall-Sundrum
scenario \cite{Randall:1999ee}.

\section{GENERAL CONSIDERATIONS}
\label{sec:general}

That new interactions unique to the third generation can lead to a
strong correlation between the non-universal $\Zbb$ and the flavor
non-diagonal $\Zdjdi$ vertices has been shown in
\cite{Haisch:2007ia}. Whereas the former structure is probed by the
ratio of the $Z$-boson decay width into bottom quarks and the total
hadronic width, $\Rb$, the bottom quark asymmetry parameter, $\Ab$,
and the forward-backward asymmetry for bottom quarks, $\AFB$, the
latter ones appear in many $K$- and $B$-decays.

In the effective field theory framework of MFV
\cite{D'Ambrosio:2002ex}, it is easy to see how the $\Zbb$ and
$\Zdjdi$ operators are linked together. The only relevant
dimension-six contributions compatible with the flavor group of MFV
stem from the $SU(2) \times U(1)$ invariant operators
\beq \label{eq:zoperators} 
{\cal O}_1 = i \left( {\bar Q}_L Y_U Y_U^\dagger \gamma_\mu Q_L
\right) \phi^\dagger D^\mu \phi \, , \qquad {\cal O}_2 = i \left(
  {\bar Q}_L Y_U Y_U^\dagger \tau^a \gamma_\mu Q_L\right) \phi^\dagger
\tau^a D^\mu \phi \, ,
\eeq 
that are built out of the quark doublets $Q_L$, the Higgs field
$\phi$, the up-type Yukawa matrices $Y_U$, and the $SU(2)$ generators
$\tau^a$. After EW symmetry breaking, ${\cal O}_{1, 2}$ are
responsible for both the effective $\Zbb$ and $\Zdjdi$ vertex. Since
all up-type quark Yukawa couplings except the one of the top, $y_t$,
are small, one has $(Y_U Y_U^\dagger)_{ji} \sim y_t^2 V_{tj}^\ast
V_{ti}$ and only this contribution matters in \Eq{eq:zoperators}.

Within the SM the Feynman diagrams responsible for the enhanced top
correction to the $\Zbb$ coupling also generate the $\Zdjdi$
operators. In fact, in the limit of infinite top quark mass the
corresponding amplitudes are up to Cabibbo-Kobayashi-Maskawa (CKM)
factors identical. Yet there is a important difference between
them. While for the physical $\Ztobb$ decay the diagrams are evaluated
on-shell, in the case of the low-energy $\Ztodjdi$ transitions the
amplitudes are Taylor-expanded up to zeroth order in the external
momenta. As far as the momentum of the $Z$-boson is concerned the two
cases correspond to $q^2 = \MZ^2$ and $q^2 = 0$.

\begin{figure}[t!]
\begin{center}
\mbox{
\begin{picture}(245,145)(0,0)
\put(0,0){\makebox{\hspace{+0mm} \scalebox{0.775}{\hspace{2mm}\includegraphics{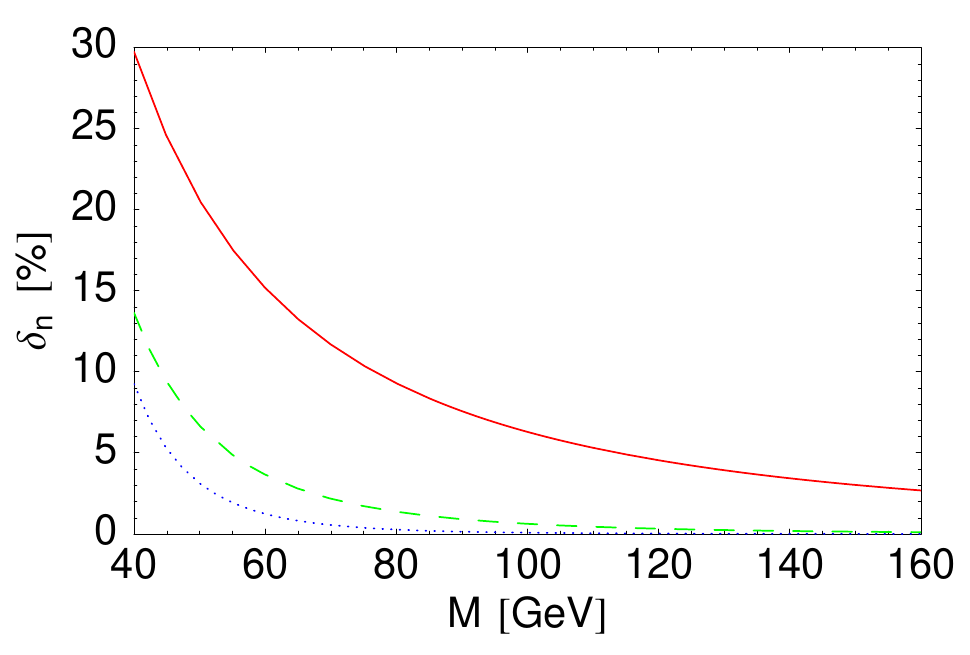}}}}
\put(95,60){\makebox{\hspace{+0mm} \scalebox{0.375}{\includegraphics{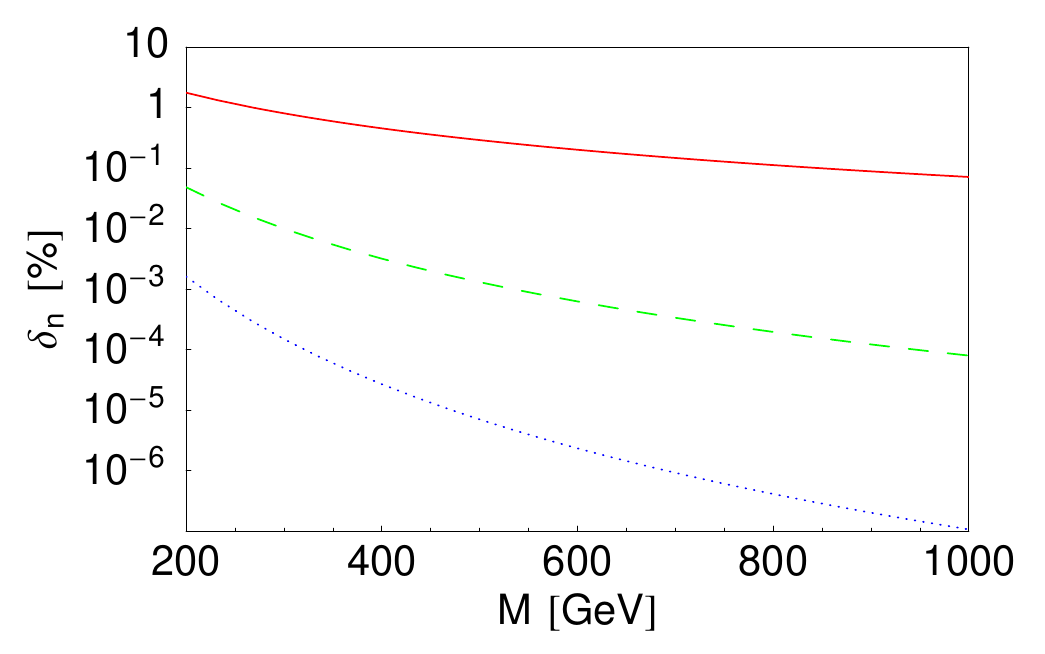}}}}
\end{picture}
}
\qquad \quad  
\mbox{
\scalebox{0.75}{\includegraphics{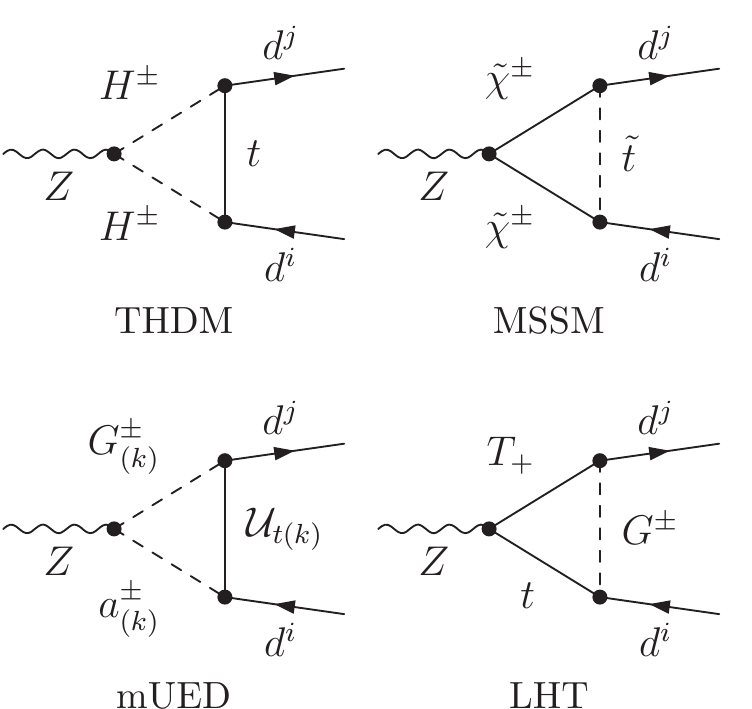}}
}
\end{center}
\vspace{-2mm}
\caption{Left: Relative deviations $\delta_n$ as a function of
  $M$. The solid, dashed, and dotted curve correspond to $n = 1,2,$
  and $3$, respectively. Right: Examples of one-loop vertex diagrams
  that result in a non-universal correction to the $\Ztodjdi$
  transition in assorted BSM scenarios with CMFV. See text for
  details.}
\label{fig:one}
\end{figure}

The general features of the small momentum expansion of the one-loop
$\Ztobb$ vertex can be nicely illustrated with the following simple
but educated example. Consider the scalar integral
\beq \label{eq:c0}
C_0 = \f{m_3^2}{i \pi^2} \int \! \f{d^4 l}{D_1 D_2 D_3} \, ,
\qquad D_i = (l + p_i)^2 - m_i^2 \, ,
\eeq
with $p_3 = 0$. In the limit of vanishing bottom quark mass one has
for the corresponding momenta $p^2_1 = p^2_2 = 0$. The small momentum
expansion of the scalar integral $C_0$ then takes the form
\beq \label{eq:sme}
C_0 = \sum_{n = 0}^\infty a_n \left ( \f{q^2}{m_3^2} \right )^n \, ,
\eeq
with $q^2 = (p_1 - p_2)^2 = -2 \hspace{0.2mm} p_1 \! \cdot \!
p_2$. Analytic expressions for the expansion coefficients $a_n$ have
been given in \cite{Haisch:2007ia}. Here we confine ourselves to the
simplified case $m_1 = m_2 = M$ and $m_3 = \mt$. We define
\beq \label{eq:deltan}
\delta_n = a_n \left ( \f{\MZ^2}{\mt^2} \right )^n \left ( \sum_{l =
    0}^{n - 1} a_l \left ( \f{\MZ^2}{\mt^2} \right )^l \right )^{-1}
\, ,
\eeq
for $n \geq 1$. The $M$-dependence of the relative deviations
$\delta_n$ is displayed on the left in \Fig{fig:one}. We see that
while for $M \lesssim 50 \, \GeV$ higher order terms in the small
momentum expansion have to be included in order to approximate the
exact on-shell result accurately, in the case of $M \gtrsim 150 \,
\GeV$ the first correction is small and higher order terms are
negligible. For the two reference scales $M = \{ 80, 250 \} \, \GeV$
one finds for the first three relative deviations $\delta_n$
numerically $+9.3 \%$, $+1.4 \%$, and $+0.3 \%$, and $+1.1 \%$, $+0.02
\%$, $+0.00004 \%$, respectively.

Of course the two reference points $M = \{ 80, 250 \} \, \GeV$ have
been picked for a reason. While the former describes the situation in
the SM, {\it i.e.}, the exchange of two pseudo Goldstone bosons and a
top quark, the latter presents a possible BSM contribution involving
besides the top, two heavy scalars. The above example indicates that
the differences between the $\Zbb$ form factor evaluated on-shell and
at zero external momenta are in general much less pronounced in models
with new heavy degrees of freedom than in the SM. Given that this
difference amounts to around $-30 \%$ in the SM, it is suggestive to
assume that the scaling of BSM contributions to the non-universal
$\Zbb$ vertex is in general under $\pm 10 \%$. This model-independent
conclusion is well supported by the results of the calculations of the
one-loop $\Zbb$ vertices in popular CMFV models presented in
\cite{Haisch:2007ia}.

\section{MODEL CALCULATIONS}
\label{sec:calculations}

The above considerations can be corroborated in another, yet
model-dependent way by explicitly calculating the difference between
the value of the $\Zdjdi$ vertex form factor evaluated on-shell and at
zero external momenta. In \cite{Haisch:2007ia} this has been done in
four of the most popular, consistent, and phenomenologically viable
scenarios of CMFV, {\it i.e.}, the two-Higgs-doublet model (THDM) type
I and II, the minimal supersymmetric SM (MSSM) with MFV, all for small
$\tan \beta$, the minimal universal extra dimension (mUED) model
\cite{Appelquist:2000nn}, and the littlest Higgs model
\cite{Arkani-Hamed:2002qy} with $T$-parity (LHT) \cite{tparity} and
degenerate mirror fermions \cite{Low:2004xc}. Examples of diagrams
that contribute to the $\Ztodjdi$ transition in these models can be
seen on the right of \Fig{fig:one}. In the following we will briefly
summarize the most important findings of \cite{Haisch:2007ia}.

In the limit of vanishing bottom quark mass, possible non-universal
BSM contributions to the renormalized off-shell $\Zdjdi$ vertex can be
written as
\beq \label{eq:zdjdi}
\Gamma_{ji}^{\rm BSM} = \f{\GF}{\sqrt{2}} \f{e}{\pi^2} \MZ^2
\f{\cw}{\sw} V_{tj}^\ast V_{ti} C_{\rm BSM} (q^2) \bar{d^j}_\sL
\gamma_\mu {d^i}_\sL Z^\mu \, ,   
\eeq
where $i = j = b$ and $i \neq j$ in the flavor diagonal and
off-diagonal case. $\GF$, $e$, $\sw$, and $\cw$ denote the Fermi
constant, the electromagnetic coupling constant, the sine and cosine
of the weak mixing angle, respectively, while $V_{ij}$ are the
corresponding CKM matrix elements.

As a measure of the relative difference between the complex valued
form factor $C_{\rm BSM} (q^2)$ evaluated on-shell and at zero
momentum we introduce
\beq \label{eq:dcnp}
\delta C_{\rm BSM} = 1 - \f{\re \, C_{\rm BSM} (q^2 = 0)}{\re \, C_{\rm
    BSM} (q^2 = \MZ^2)} \, .
\eeq  

\begin{figure}[!t]
\begin{center}
\scalebox{0.5}{\includegraphics{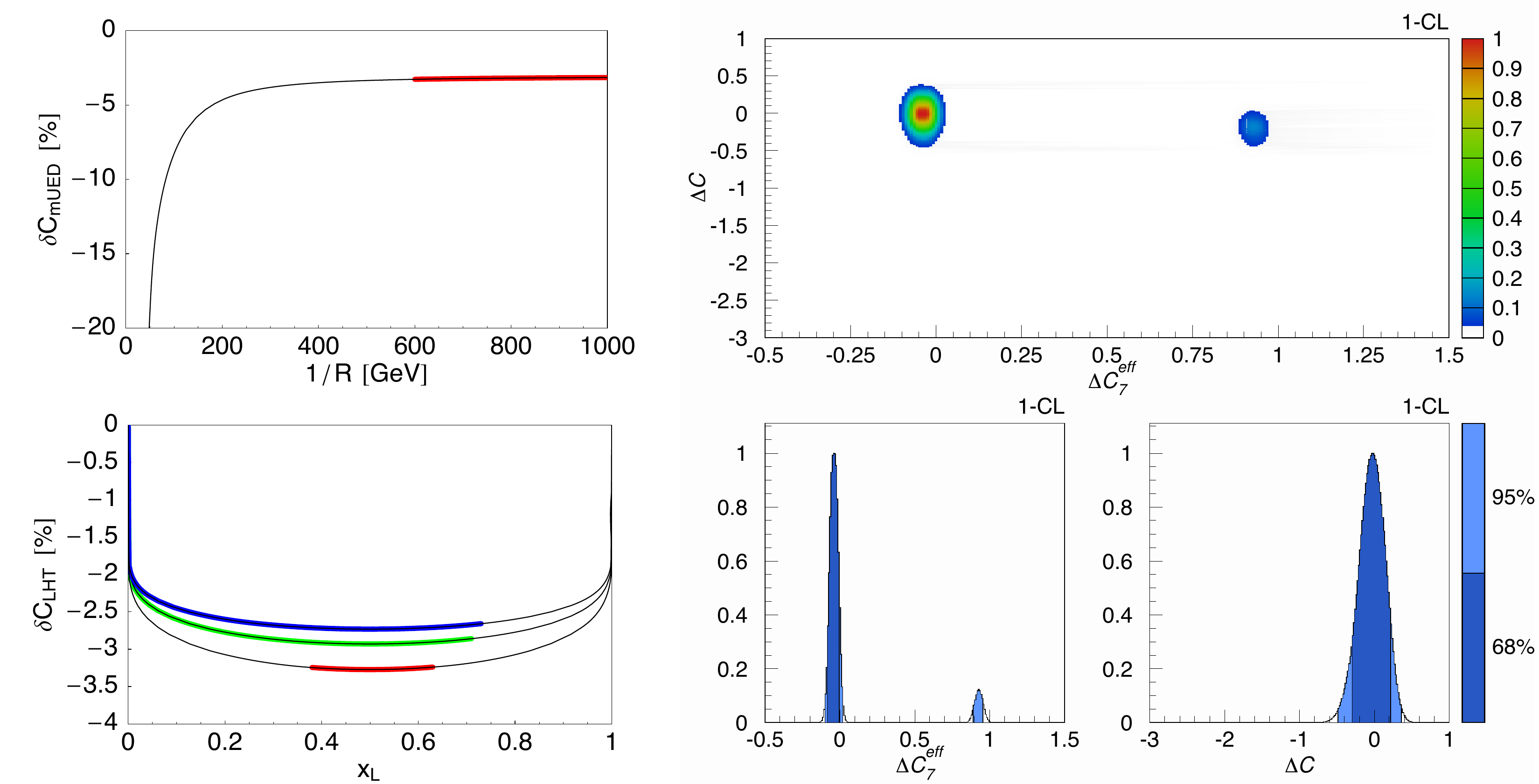}}
\end{center}
\vspace{-6mm}
\caption{Left: Relative difference $\delta C_{\rm BSM}$ in the mUED
  and the LHT model as a function of $1/R$ and $x_L$. In the case of
  the LHT model the shown curves correspond, from bottom to top, to
  the values $f = 1, 1.5$, and $2 \, \TeV$ of the symmetry breaking
  scale. Right: Constraints on $\Delta C_7^{\rm eff}$ and $\Delta C$
  within CMFV that follow from a combination of the $\Ztobb$ pseudo
  observables with the measurements of $\BXsga$ and $\BXsll$. The
  colors encode the frequentist $1 - {\rm CL}$ level and the
  corresponding $68 \%$ and $95 \%$ probability regions as indicated
  by the bars on the right side of the panels. See text for details.}
\label{fig:two}
\end{figure}

The dependence of $\delta C_{\rm BSM}$ on the compactification scale
$1/R$ of the mUED model and $x_L$, which parametrizes the mass of the
heavy top $T_+$ in the LHT scenario, is illustrated in the two plots
on the left-hand side in \Fig{fig:two}. The allowed parameter regions
after applying the $\bar{B} \to X_s \gamma$ constraint in the case of
the mUED model \cite{Haisch:2007vb} and electroweak precision
measurements in the case of the LHT scenario \cite{Hubisz:2005tx} are
indicated by the colored (grayish) bands.

In the THDMs, the mUED, and the CMFV version of the LHT model the
maximal allowed suppressions of $\re \, C_{\rm BSM} (q^2 = \MZ^2)$
with respect to $\re \, C_{\rm BSM} (q^2 = 0)$ amounts to less than $2
\%$, $5 \%$, and $4 \%$, respectively. This feature confirms the
general argument presented in the last section. The situation is less
favorable in the case of the CMFV MSSM, since $\delta C_{\rm MSSM}$
frequently turns out to be larger than one would expected on the basis
of the model-independent considerations if the masses of the lighter
chargino and stop both lie in the hundred $\GeV$ range. However, the
large deviation $\delta C_{\rm MSSM}$ are ultimately no cause of
concern, because $|\re \, C_{\rm MSSM} (q^2 = 0)/\re \, C_{\rm SM}
(q^2 = 0)|$ itself is always below $10 \%$. In consequence, the
model-independent bounds on the BSM contribution to the universal
$Z$-penguin function that will be derived in the next section do hold
in the case of the CMFV MSSM. More details on the phenomenological
analysis of $\delta C_{\rm BSM}$ in the THDMs, the CMFV MSSM, the
mUED, and the LHT model including the analytic expressions for the
form factors $C_{\rm BSM} (q^2)$ can be found in \cite{Haisch:2007ia}.

\section{NUMERICAL ANALYSIS}
\label{sec:numerics}

The BSM contribution $\Delta C = {\rm Re} \hspace{0.5mm} C(q^2 = 0) -
{\rm Re} \hspace{0.5mm} C_{\rm SM}(q^2 = 0)$ can be extracted in a
model-independent fashion from a global fit to the pseudo observables
$\Rb$, $\Ab$, and $\AFB$ and the measured $\BXsga$ and $\BXsll$
branching ratios. Neglecting contributions from EW boxes these bounds
read \cite{Haisch:2007ia}
\beq \label{eq:dcsb0}
\Delta C = -0.026 \pm 0.264 \;\; (68 \% \, {\rm CL}) \, , \qquad 
\Delta C = [-0.483, 0.368] \;\; (95 \% \, {\rm CL}) \, . 
\eeq 
These numbers imply that large negative contributions that would
reverse the sign of the SM $Z$-penguin amplitude are highly disfavored
in CMFV scenarios due to the strong constraint from $\Rb$
\cite{Haisch:2007ia}. Interestingly, such a conclusion cannot be drawn
by considering only flavor constraints \cite{Bobeth:2005ck}, since at
present, a combination of $\BRga$, $\BRll$, and $\BRKp$ does not allow
to distinguish the SM solution $\Delta C = 0$ from the wrong-sign case
$\Delta C \approx -2$. The constraints on $\Delta C$ within CMFV
following from the simultaneous use of $\Rb$, $\Ab$, $\AFB$, $\BRga$,
and $\BRll$ can be seen on the right-hand side of \Fig{fig:two}.

One can also infer from this figure that two regions, resembling the
two possible signs of the amplitude ${\cal A} (\btosgamma) \propto
C_7^{\rm eff} (\mb)$, satisfy all existing experimental bounds. The
best fit value for $\Delta C_7^{\rm eff} = C_7^{\rm eff} (\mb) - C_{7
  \, {\rm SM}}^{\rm eff} (\mb)$ is very close to the SM point residing
in the origin, while the wrong-sign solution located on the right is
highly disfavored, as it corresponds to a $\BRll$ value considerably
higher than the measurements \cite{Gambino:2004mv}. The corresponding
limits are \cite{Haisch:2007ia}
\beq \label{eq:dc7effsb0}
\Delta C_7^{\rm eff} = -0.039 \pm 0.043 \;\; (68 \% \, {\rm CL})
\, , \qquad 
\Delta C_7^{\rm eff} = [-0.104, 0.026] \, \cup \, [0.890, 0.968] 
\;\;  (95 \% \, {\rm CL}) \, . 
\eeq
Similar bounds have been presented previously in
\cite{Bobeth:2005ck}. Notice that since the SM prediction of $\BRga$
\cite{bsg} is now lower than the experimental world average by $1.2
\mysigma$, extensions of the SM that predict a suppression of the
$\btosgamma$ amplitude are strongly constrained. In particular, even
the SM point $\Delta C_7^{\rm eff} = 0$ is almost disfavored at $68 \%
\, {\rm CL}$ by the global fit. 

The stringent bound on the BSM contribution $\Delta C$ given in
\Eq{eq:dcsb0} translates into tight two-sided limits for the branching
ratios of all $Z$-penguin dominated flavor-changing $K$- and
$B$-decays as shown in \Tab{tab:brs}. A strong violation of any of the
bounds by future measurements will imply a failure of the CMFV
hypothesis, signaling either the presence of new effective operators
and/or new flavor and CP violation. A way to evade the given limits is
the presence of sizable corrections $\delta C_{\rm BSM}$ and/or box
contributions. While these possibilities cannot be fully excluded,
general arguments and explicit calculations indicate that they are
both difficult to realize in the CMFV framework.

\begin{widetext}
\begin{center}
\begin{table}[!t]
\begin{center}
\begin{tabular}{c@{\hspace{2.5mm}}cccc}
\hline \hline \\[-4.5mm]
Observable & CMFV ($95 \% \, {\rm CL}$) & \hspace{0mm} SM ($68 \% \,
{\rm CL}$) & \hspace{0mm} SM ($95 \% \, {\rm CL}$) & \hspace{0mm}
Experiment \\[0.5mm] 
\hline 
$\BRKp \times 10^{11}$ & $[4.29, 10.72]$ & \hspace{0mm} $7.15 \pm 1.28$ &  
\hspace{0mm} $[5.40, 9.11]$ & \hspace{0mm} $\left ( 17.3^{+11.5}_{-10.5}
\right )$ \cite{Artamonov:2008qb} \\
$\BRKL \times 10^{11}$ & $[1.55, 4.38]$ & \hspace{0mm} $2.79 \pm 0.31$
& \hspace{0mm} $[2.21, 3.45]$ & \hspace{0mm} $< 6.7 \times 10^3 \;\;
(90 \% \, \text{CL})$ \cite{Ahn:2007cd} \\
$\BRKm \times 10^9$ & $[0.30, 1.22]$ & \hspace{0mm} $0.70 \pm 0.11$ &
\hspace{0mm} $[0.54, 0.88]$ & \hspace{0mm} -- \\
$\BRXd \times 10^6$ & $[0.77, 2.00]$ & \hspace{0mm} $1.34 \pm 0.05$ &
\hspace{0mm} $[1.24, 1.45]$ & \hspace{0mm} -- \\
$\BRXs \times 10^5$ & $[1.88, 4.86]$ & \hspace{0mm} $3.27 \pm 0.11$ &
\hspace{0mm} $[3.06, 3.48]$ & \hspace{0mm} $< 64 \;\; (90 \% \,
\text{CL})$ \cite{Barate:2000rc} \\ 
$\BRBd \times 10^{10}$ & $[0.36, 2.03]$ & \hspace{0mm} $1.06 \pm 0.16$ &
\hspace{0mm} $[0.87, 1.27]$ & \hspace{0mm} $< 1.8 \times 10^2 \;\; (95
\% \, \text{CL})$ \cite{Aaltonen:2007kv} \\  
$\BRBs \times 10^9$ & $[1.17, 6.67]$ & \hspace{0mm} $3.51 \pm 0.50$ &
\hspace{0mm} $[2.92, 4.13]$ & \hspace{0mm} $< 5.8 \times 10^1
\;\; (95 \% \, \text{CL})$ \cite{Aaltonen:2007kv} \\[1mm] 
\hline \hline
\end{tabular}
\end{center}
\caption{Bounds for various rare decays in CMFV models at $95 \%$
  probability, the corresponding values in the SM at $68 \%$ and $95
  \% \, {\rm CL}$, and the available experimental information. See
  text for details.}  
\label{tab:brs}
\end{table}
\end{center}
\end{widetext}

\section{CONCLUSIONS}
\label{sec:conclusions}

We have emphasized that large contributions to the universal Inami-Lim
function $C$ in constrained minimal flavor violation that would
reverse the sign of the standard $Z$-penguin amplitude are excluded by
the existing measurements of the $\Ztobb$ pseudo observables performed
at LEP and SLC. This underscores the outstanding role of electroweak
precision tests in guiding us toward the right theory and immediately
raises the question: what else can flavor physics learn from the
high-energy frontier?

\begin{center}
\end{center}

\end{document}